\documentclass[12pt]{article}
\usepackage{amsmath}
\usepackage{amssymb}
\usepackage{graphicx}
\usepackage[cp1251]{inputenc}
\usepackage[russian]{babel}
\newcommand{\be}{\begin{equation}}
\newcommand{\ee}{\end{equation}}
\newcommand{\ba}{\begin{eqnarray}}
\newcommand{\ea}{\end{eqnarray}}

\begin{document}
\vspace{1cm}

\begin{center}
{\bf Magnetic properties of nanosized diluted magnetic
semiconductors\\ with band splitting}
\end{center}
  \vspace{1cm}

\centerline{E.Z. Meilikhov\footnote{$^)$ e-mail:
meilikhov@imp.kiae.ru}$^)$, R.M. Farzetdinova}
\medskip
\centerline{\small\it RRC “Kurchatov Institute”, 123182 Moscow,
Russia}
  \vspace{1cm}

  \centerline{
\begin{tabular}{p{15cm}}
\footnotesize\qquad The continual model of the nonuniform magnetism in thin
films and wires of a diluted magnetic semiconductor is considered with taking
into account the finite spin polarization of carriers responsible for the
indirect interaction of magnetic impurities (e.g. via RKKY mechanism). Spatial
distributions (across the film thickness or the wire radius) of the
magnetizaton and carrier concentrations of different spin orientations, as well
as the temperature dependence of the average magnetization are determined as
the solution of the nonlinear integral equation.
\end{tabular}
} \vspace{0.5cm}

The indirect interaction of magnetic impurities, e.g. of the
Ruderman-Kittel-Kasuya-Yosida (RKKY) type, is believed to be one of the basic
mechanisms of magnetic ordering in systems with free carriers of a high
concentration (metals and degenerate semiconductors). Since the most of
potentially interesting electron devices are of nanometer sizes, it is of
interest to see how magnetic properties of the relevant systems depend on their
\emph{finite} sizes~\cite{1,2}. The typical example is the thin film of the
diluted magnetic semiconductor (e.g., Ga$_{1-x}$Mn$_x$As) with the thickness on
the order of some tens of the lattice periods which are considered in the
present paper. In such systems, the mobility of charge carriers (holes) is low
(on the order of 10 cm$^2$/V$\cdot$s~\cite{3}) and, hence, their mean free path
is also short. Then the collisional broadening of the hole energy levels is so
high that the system appears to be effectively three dimensional one, and there
is no need to take into account effects of the size quantization~\cite{4}.

The magnetization of a thin enough film of the diluted magnetic semiconductor
(whose thickness $L$ is comparable with the characteristic length $\ell$ of the
magnetic impurities' indirect interaction, see below) could be essentially
nonuniform, at least,  in the direction perpendicular to the film plane. It is
convenient to characterize such a nonuniform magnetization, depending on
coordinates, by the reduced local magnetization $j({\bf r})\equiv M({\bf
r})/M_s$ ($M_s$ is the saturation magnetization). Non-zero local magnetization
($0\leq j\leq 1$ ) of Mn atoms with the spin $S_{\rm Mn}=5/2$ leads to the
nonuniform local spin polarization of holes reflecting in the fact that the
local concentration $p^-({\bf r})$ of holes with spins, antiparallel to the
local magnetization, exceeds the concentration $p^+({\bf r})$ of holes with the
opposite spin direction. At that, the local hole polarization $\xi({\bf
r})=[p^-({\bf r})-p^+({\bf r})]/p({\bf r})$ does not equal zero (\,$p({\bf
r})=p^-({\bf r})+p^+({\bf r})$ is the total hole concentration). Formally, the
hole polarization could be  related with the effective spin-depending magnetic
potential~\cite{5}
 \be\label{1}
V_{\rm mag}({\bf r})=xN_0a^3J_{pd}\,\sigma S_{\rm Mn}j({\bf r}),
 \ee
influencing on the holes. Here $N_0\approx8\cdot10^{21}$ cm$^{-3}$ is the
concentration of Ga sites in the GaAs lattice, $x$ is the fraction of such
sites populated by Mn atoms, $a$ being the lattice period, and  $\sigma=\pm
1/2$ is the hole spin, $J_{pd}=1.2$ eV is the exchange energy between mobile
holes and localized $d$-electrons of Mn atoms~\cite{6}. According to (\ref{1}),
major  holes (with the preferred spin orientation) are accumulated in regions
of the high magnetization, and, contrary,  minor ones are pushed out in regions
with the low magnetization.

Such a spin separation of carriers in the space could be, in
principle, accompanied by modifying the spatial distribution of
holes resulting in the violation of the local charge neutrality. To
see if that possible in the considered case, notice that the
characteristic displacement of the hole charge “centroid” over a
distance $\delta L$  leads to the electric potential equal to
 \be\label{2}
  V_{\rm
el}\sim \frac{4\pi p e^2}{\kappa}(\delta L)^2,
 \ee
where $\kappa\sim10$ is the semiconductor  dielectric constant.  From
(\ref{1}), (\ref{2})  one finds $V_{\rm el}/|V_{\rm mag}|\sim e^2L^2/J_{pd}$.
Herefrom, it follows $V_{\rm el}/|V_{\rm mag}|\gg1$ at $\delta L\gtrsim 10$
\AA. This suggest that for films of the thickness $L\gg10$ \AA\, one could
neglect the violation of the electric neutrality and hold $p({\bf r})=p^-({\bf
r})+p^+({\bf r})={\rm Const}$. In the present paper, we shall not go beyond
considering such a case.

Let the Fermi energy of charge carriers in the absence of the
magnetization ($j=0$) be $\varepsilon_F$. Magnetic potential
(\ref{1}), arising with the origination of the magnetization, leads
to splitting the hole band into two spin sub-bands with effective
Fermi energies (measured from bottoms of those sub-bands)
$\varepsilon_F+V_{\rm mag}({\bf r})$ and $\varepsilon_F-V_{\rm
mag}({\bf r})$ which correspond to following Fermi momenta
 \be\label{3}
k_F^{\pm}({\bf r})=k_F\left(1\pm\frac{V_{\rm mag}({\bf
r)}}{\varepsilon_F}\right)^{1/2},
 \ee
where $k_F=(2m^*\varepsilon_F/\hbar^2)^{1/2}$, $m^*\approx0.5m_0$ is
the effective hole mass (at $V_{\rm mag}({\bf r})/\varepsilon_F>1$,
${k_F^-({\bf r})=0}$).

Carrier concentrations in the sub-bands are determined by the set of
simple equations
 \be\label{40}
p^-({\bf r})\propto[\varepsilon_F+V_{\rm mag}({\bf
r})\,]^{3/2},\quad p^+({\bf r})\propto[\varepsilon_F-V_{\rm
mag}({\bf r})\,]^{3/2},\quad p^-({\bf r})+p^+({\bf r})=p,
 \ee
from which it follows
 \be\label{50}
 p^-({\bf r})=\frac{p}{2}\,[1+\xi({\bf r})],\quad p^+({\bf
r})=\frac{p}{2}\,[1-\xi({\bf r})],
 \ee
where
 \be\label{60} \xi({\bf r})=\frac{u({\bf r})-1}{u({\bf r})+1}
 \ee
is the hole spin polarization,
 $u({\bf r})=\{[1+V_{\rm mag}({\bf r})/\varepsilon_F]/[1-V_{\rm
mag}({\bf r})/\varepsilon_F]\}^{3/2}$ (if $V_{\rm mag}({\bf
r})/\varepsilon_F>1$, then $p^-=p$, $p^+=0$, $\xi=1$).

With the uniform magnetization ($j={\rm Const}$), momenta $k_F^\pm $
do not depend on coordinates and the expression for the energy
$w(\rho)$ of the indirect interaction of two magnetic atoms,
separated by a distance  $\rho$ from each other, taking into account
the spin splitting of the hole band could be written in the form
 \be\label{4}
w(\rho)=-\frac{1}{2}I_0 \Phi(\rho)\exp(-\rho/\ell),
 \ee
where, as an example, the RKKY range function reads~\cite{7}
 \be\label{5}
 \Phi(\rho)=\left(\frac{a}{\rho}\right)^4
[\theta^+(\rho)\cos\theta^+(\rho)-\sin\theta^+(\rho)+\theta^-(\rho)\cos\theta^-(\rho)-\sin\theta^-(\rho)],
 \ee
 \be\label{6}
I_0=\frac{1}{32\pi^3}\left(\frac{ma^2}{\hbar^2}J_{pd}^2\right),\quad
\theta^\pm(\rho)=2k_F^\pm \rho.
 \ee
Exponential factor in (\ref{4}) allows for the finite length $\ell$
of the hole spin relaxation~\cite{8} (in the simplest case,
coinciding with their mean free path).

With increasing magnetic potential (\ref{1}), the intensity of the interaction
(\ref{4}) is changed. One could judge that from the behaviour of the function
$\Phi(\rho)$ at distances close to the mean distance $\bar\rho$ between
impurities. Spatial dependencies of the RKKY-function $\Phi(\rho)$ for various
values of the ratio $V_{\rm mag}/\varepsilon_F$ are shown in Fig. 1. It is
clear that in the inter-impurity distance range of $\bar\rho=1.5a-2a$,
corresponding to the impurity concentration $x=0.05-0.1$, the interaction
energy (at a fixed $\varepsilon_F$ value) decreases noticeably with increasing
$V_{\rm mag}$. What that means is the spin splitting of the hole band has to
result, in the end, in decreasing the system magnetization.

The generalization of the function (\ref{5}) over the case of the
non-uniform magnetization could be done with replacing phases
$\theta^\pm(\rho)$ by their average values
 \be\label{7}
\theta^*_\pm(\rho)=2\int\limits_0^\rho k_F^\pm
(s)ds=2k_F\int\limits_0^\rho \sqrt{1\pm Aj(s)}ds,
 \ee
where $A=xN_0a^3J_{pd}\sigma S_{\rm Mn}/\varepsilon_F$, and the
integration is performed along the line connecting the impurities.
Then the function $\Phi(\rho)$ becomes the functional of the
spatially non-uniform magne\-ti\-za\-tion:
 \ba\label{8}
\Phi(\rho)\exp(-\rho/\ell)\to\hat F [Aj({\bf r}), \rho]
\equiv\hspace{90pt}\nonumber\\
\label{8}\equiv\left(\frac{a}{\rho}\right)^4
[\theta_+^*(\rho)\cos\theta_+^*(\rho)-\sin\theta_+^*
(\rho)+\theta_-^*(\rho)\cos\theta_-^*(\rho)-\sin\theta_-^*(\rho)]e^{-\rho/\ell}.
 \ea

Under low magnetization ($j\ll 1$), the magnetic potential (\ref{1})
is also low, and the expression (\ref{8}) goes to the standard form.

The finite carrier polarization is important in the only case when the magnetic
potential and Fermi energy are comparable, that is at $Aj\sim1$. Let us
estimate the parameter $A$ for the magnetic semiconductor Ga$_{1-x}$Mn$_x$As.
Due to the compensation, the hole concentration $p$ is always lower than the
concentration $xN_0$ of Mn atoms (acceptors) positioned in Ga-sites.
Nevertheless, $p/(xN_0)\gtrsim 0.3$ at $x=0.05$~\cite{9}. The estimate for that
case leads to $A\gtrsim 1$. If, in addition, the local magnetization in some
regions of the system is close to the saturation ($j\sim 1$), then $Aj\sim 1$
and taking account of the finite carrier polarization could be important.

To test the validity of that conclusion, we have carried out calculating the
spatial distribution of the magnetization in a thin magnetic semiconductor film
where the magnetization is parallel to the film surface. In doing so, we have
employed the approach elaborated early~\cite{4}.

The energy $W_{\rm RKKY}$ of the indirect interaction of a given spin $S_i$
with its surraundings is defined by the relation  $W_{\rm
RKKY}=\sum_{j=1}^\infty w(\rho_{ij})$. The distance $\rho_{ij}$ could not be
less than the distance $r_{\rm min}$ between the neighbor lattice sites,
accessible for magnetic impurities (for the diluted semiconductor
Ga$_{1-x}$Mn$_x$As, $r_{\rm min}=a/\sqrt{2}$ where $a\approx5$\AA\, being the
lattice constant). In the continual approximation, the sum could be replaced by
the integral
 \be\label{9}
W_{\rm RKKY}(\textbf{r})=-\frac{1}{2}n_\mu I_0\int \hat F [Aj({\bf
r}'), |{\bf r}-{\bf r'}|]j(\textbf{r}')d^3\textbf{r}',
 \ee
where $n_\mu$ is the concentration of magnetic Mn-ions, and the integration is
expanded over the volume occupied by impurities.

Contrary to the infinite system, the value of that integral depends
on the system geometry and the position of a chosen point. Below, we
consider the magnetic interaction in the film of the thickness $L$
(in the $z$-axis direction) which is so thin that the local
magnetization $j(h)$ within the film is always parallel to the
surface and depends on the given point distance $h$ from the median
film plane only\footnote{That means we consider systems with weak
enough surface magnetic anisotropy~\cite{10}.}. In that case, it
follows from (\ref{9})
 \be\label{10}
 W_{\rm RKKY}(h)=-\pi
n_\mu I_0\int\limits_{z=-L/2}^{L/2}\left(\,\,\int\limits_{\rho_{\rm
min}(z,h)}^{\infty}\hat F [Aj(z),\rho]\rho\, d\rho\right)j(z) dz,
 \ee
where
 \be\label{11}
  \rho_{\rm min}(z,h)={\rm Max}[\,|\,h-z|,\, r_{\rm min}\,].
\ee

The relation (\ref{10}), defining the \emph{local} energy $W_{\rm
RKKY}(h)$ in all film points, is \emph{non-local} one: that energy
is the functional of the local magnetization $j(\rho)$ and specified
by \emph{all} film points.

The self-consisted equation $j(h)=\tanh[W_{\rm RKKY}(h)/kT]$,
determining the local magnetization, could be now written as follows
 \be\label{12}
j(h)=\tanh\left[-\frac{\pi n_\mu I_0
}{kT}\int\limits_{z=-L/2}^{L/2}\left(\,\,\int\limits_{\rho_{\rm
min}(z,h)}^{\infty}\hat F [Aj(z),\rho]\rho\, d\rho\right)j(z)
dz\right],
 \ee
where phases appearing in (\ref{8}) are defined by relations
 \be\label{13}
\theta_\pm^*(\rho,z,h)=2k_F\hspace{-15pt}\int\limits_{\rho_{\rm
min}(z,h)}^\rho\hspace{-15pt}\sqrt{1\pm
Aj[h-(\rho'/\rho)(h-z)]}\,\,d\rho'.
 \ee

Introducing the designation
 \be\label{14}
-\pi n_\mu\hspace{-15pt}\int\limits_{\,\,\rho_{\rm
min}(z,\,h)}^{\infty}\hspace{-10pt}\hat F [Aj(z),\rho]\rho
d\rho={K}(z,h)
 \ee
for the “internal” integral in Eq.(\ref{12}), one could write it in the form
 \be\label{15}
j(h)=\tanh\left[\frac{1}{\tau}\hspace{-10pt}\int\limits_{z=-L/2}^{L/2}\!\!\!{
K}(z,h)\,j(z) dz\right],
 \ee
where $\tau=kT/I_0$ is the reduced temperature.

The \emph{non-linear} integral equation (\ref{15}) determines the spatial
distribution of the magnetization in the considered system at a given
temperature. It differs from the equation considered early\cite{4} in that its
kernel is the functional of the non-uniform magnetization (due to the finite
carrier spin polarization). The solution of the obtained equation has been
found, as in~\cite{4}, by the successive approximations method.

Spatial distributions of the magnetization in the film of the thickness $L=20a$
at various temperatures $\tau$ are shown in Fig.~2. The chosen value $k_Fa=1$
corresponds to the hole concentration $p\approx2\cdot10^{20}$ см$^{-3}$,
characteristic to Ga$_{1-x}$Mn$_x$As with $x=0.05$. It is clear that taking
account of the band splitting, leading to the finite carrier polarization,
decreases noticeably, as might be expected, the system magnetization.  Near the
film surface the magne\-ti\-za\-tion drops quickly, and the thickness of the
relevant subsurface regions is almost linearly increase with the temperature
(cf. Fig.~2).

The magnetization of “thick” films ($L\to\infty$) is nearly uniform (except for
“thin” near-surface regions where it is twice as low comparing to the bulk).
The effect of the band splitting evaluated for that case by the ratio  $j(A=1,
L\to\infty)/j(A=0, L\to\infty)$, is illustrated by Fig.~3, where its
temperature dependence is shown. It could be seen that at
$\tau\gtrsim0.2\tau_{\rm C}$ the effect is very notable.

It is convenient to characterize the non-uniformly magnetized film by its
average magnetization
 \be\label{101}
\langle j\rangle=\frac{1}{L}\int\limits_{-L/2}^{L/2}j(z)dz.
 \ee
Temperature dependencies of that quantity for the film of the thickness $L=20a$
are represented in Fig. 4. They allow to find the Curie temperature $T_{\rm
C}$, defined as the temperature at which $\langle j\rangle\to 0$. Near the
Curie temperature,  $j\ll 1$ and, hence, taking into account the finite spin
polarization does not change $T_{\rm C}$ value. That is immediately seen from
Fig. 4, where both dependencies $\langle j\rangle(\tau)$ (for $A=0$ and $A=1$)
give the same value $\tau_{\rm C}\approx 1$.

In the framework of the considered scheme, it is easy to find spatial
distributions of the carrier spin polarization, as well as those of major and
minor carrier concentrations. Correspon\-ding dependencies, obtained with the
help of Eqs. (\ref{50}),(\ref{60}), are shown in Fig.~4. It follows, therefrom,
that the hole polarization $\xi$ in the central part of the film is more higher
than at the periphery. That is connected with the spatial separation of
carriers with different spin polarization: major carriers concentrate away from
the film surface, and minor ones -- in peripheral regions. (In those regions,
the violation of the electric neutrality is  possible, so the relevant results
should be considered as qualitative ones.)

At $T\to T_{\rm C}$ Eq. (\ref{15}) is simplified, turning into the homogeneous
linear integral equation:
 \be\label{mft3a}
j(h)=\frac{1}{\tau_{\rm C}}\int\limits_{z=-L/2}^{L/2}\!\!\!{
K}(z,h)\,j(z) dz,
 \ee
from which one could see that the Curie temperature  $\tau_{\rm C}$
is nothing than the eigenvalue of that equation kernel. Due to the
symmetry of the kernel ($K(z,h)=K(h,z)$, see.~(\ref{14})) such an
eigenvalue is always exists ~\cite{11}. To determine it, any one of
a number of known numerical methods~\cite{12} is appropriate.
However, in practice it is more simply to find an approximate
$\tau_{\rm C}$ value by the above described approach.

In the same way, magnetic properties of a thin wire could be considered. The
generalization of the relevant integral equation~\cite{13}, for the case of the
local magnetization being everywhere oriented along the wire axis and depending
on distance $h$ to that axis only, reads
 \be\label{201}
j(h)=\tanh\left[\frac{1}{\tau}\int\limits_{0}^{R}\!{ K}(r,h)\,j(r)
dr\right],
 \ee
where $R$ is the wire radius,
 \be\label{202}
K(r,h)=-n_\mu
r\int\limits_0^{2\pi}\left(\,\,\,\int\limits_{\rho_{\rm
min}}^\infty\frac{\rho \hat
F[Aj(r),\rho]}{\sqrt{\rho^2-r^2-h^2+2rh\cos\phi}}d\rho\right)d\phi,
 \ee
$\rho_{\rm min}=\rho_{\rm min}(r,h,\phi)={\rm Max}[(r^2+h^2-2rh\cos\phi)^{1/2},
r_{\rm min}]$. In that case, phases, appearing in Eq.~(\ref{8}), are  defined
by the relations
 \be\label{203}
\theta_\pm^*=2k_F\int\limits_{0}^\rho\sqrt{1\pm
Aj[\left\{[h(1-\rho'/\rho)-r(\rho'/\rho)]^2+2rh(\rho'/\rho)(1-\rho'/\rho)(1-\cos\phi)\right\}^{1/2}]}\,\,d\rho'.
 \ee

In Fig. 5, spatial distributions of the local magnetization are displayed for
the wire of the diameter $2R=20a$ where at $\tau\sim1$ the paraxial region
occurs to be magnetized only. As well as for a film, the splitting of the
carrier band results in the essential lowering the system magnetization. E.g.,
at $\tau\sim0.1$ the magnetization in points, spaced at the distance $h=R/2$
from the axis, decreases by 2.5 times because of that effect.

The strong non-uniformity of the local magnetization results in that
temperature dependen\-cies of the average wire magnetization, shown in Fig.~6,
have unusual concave shape. In the inset of Fig.~6, spatial dependencies of the
concentrations $p^-$, $p^+$ of major and minor carriers along with their
polarization  are displayed. One could see that the noticeable hole
polarization occurs in the paraxial wire region only and (at $\tau=0.1$)
reaches $\sim 80\%$.

In conclusion, we have considered the continual model of the non-uniform
magnetism in thin films and wires of diluted magnetic semiconductor under a
finite spin polarization of carriers responsible for the indirect (RKKY)
interaction of magnetic impurities.  We succeeded in obtaining spatial
distributions of the magnetization described by the non-linear integral
equation with the help of quickly converging iterative procedure. They occurs
to be essentially non-uniform and strongly dependent on splitting the band of
carriers responsible for the indirect interaction between magnetic impurities.
Results could be used to describe properties of nanosized systems of diluted
magnetic semiconductors. \vspace{0.5cm}

\centerline{\bf Acknowledgments}

\medskip
The work is supported by Grant\# 09-02-00579 of the Russian Foundation of Basic
Researches and ISTC Grant \# G-1335.

\newpage
\renewcommand{\refname}{\centerline{\rm\small \bf References}\vspace{5mm}}

\newpage
\centerline{\bf Captions}
\bigskip
\bigskip

Fig. 1. Range functions $\Phi(\rho)$ of the inter-impurity interaction energy
(\ref{4}) at various values of the ratio  $V_{\rm
mag}/\varepsilon_F$. Accepted: $k_Fa=1$.\\

Fig. 2. Spatial distributions of the local magnetization in the film of the
thickness $L=20a$ at various temperatures  $\tau$ with ($A=1$, right panel) and
without ($A=0$, left panel) taking account of the band splitting. Accepted:
$\ell=3a$, $k_Fa=1$, $4\pi n_\mu=1$. Areas of the reduced peripheric
magnetization are marked by dashed lines.\\

Fig. 3. Temperature dependence of the band splitting effect leading to the
noticable reducing the magnetization of the  “thick” ($L\to\infty$) film.
Accepted: $\ell=3a$,
$k_Fa=1$, $4\pi n_\mu=1$.\\

Fig. 4. Temperature dependencies of the average magnetization  $\langle
j\rangle$ in the film of the thickness $L=20a$ with ($A=1$, lower curve) and
without ($A=0$, upper curve) taking account of the band splitting. Insert:
spatial distributions of the spin polarization $\xi$, and concentrations of
major ($p^-$) and minor ($p^+$) carriers  at $A=1$, $\tau=0.8$. Accepted:
$\ell=3a$, $k_Fa=1$,
$4\pi n_\mu=1$.\\

Fig. 5. Spatial distributions of the local magnetization in the wire of the
diameter  $2R=20a$ at temperatures $\tau=1$ and $\tau=0.1$ with ($A=1$, solid
curves) and without ($A=0$, dashed curves) taking account of the finite carrier
spin polarization. Accepted:  $\ell=3a$, $k_Fa=1$, $4\pi n_\mu=1$.\\

Fig. 6. Temperature dependencies of the average magnetization  $\langle
j\rangle$ in the wire of the diameter $2R=20a$ with ($A=1$, lower curve) and
without ($A=0$, upper curve) taking account of the band splitting. Insert:
spatial distributions of the spin polarization $\xi$, and concentrations of
major ($p^-$) and minor ($p^+$) carriers  at $A=1$, $\tau=1$. Accepted:
$\ell=3a$, $k_Fa=1$, $4\pi n_\mu=1$.\\

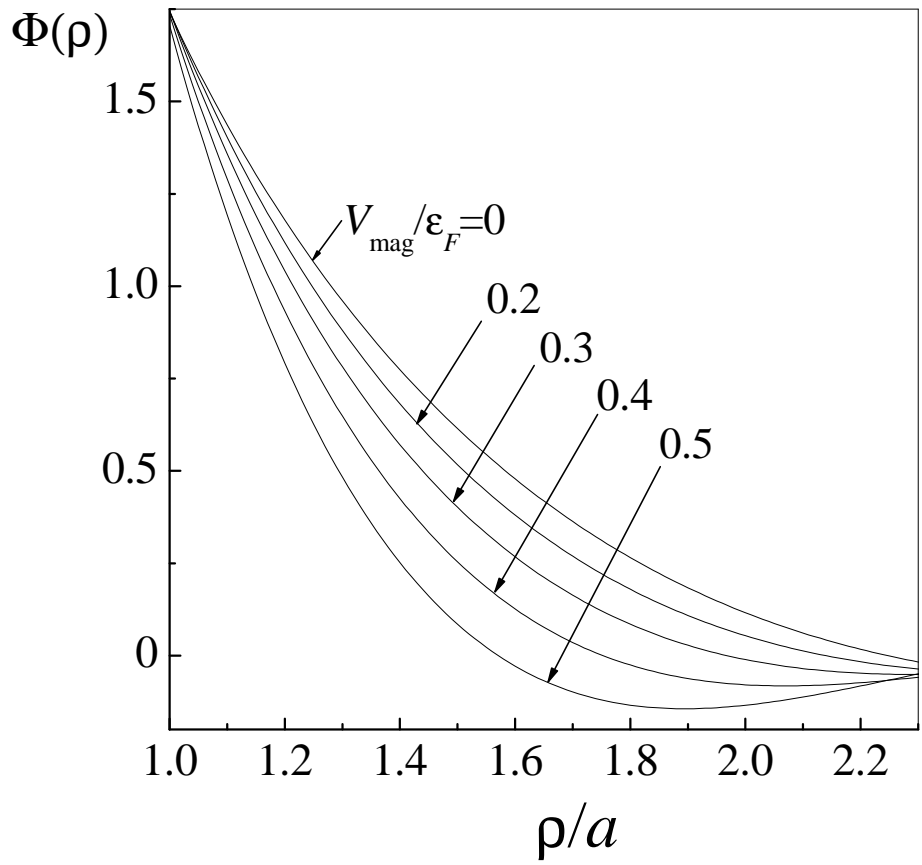
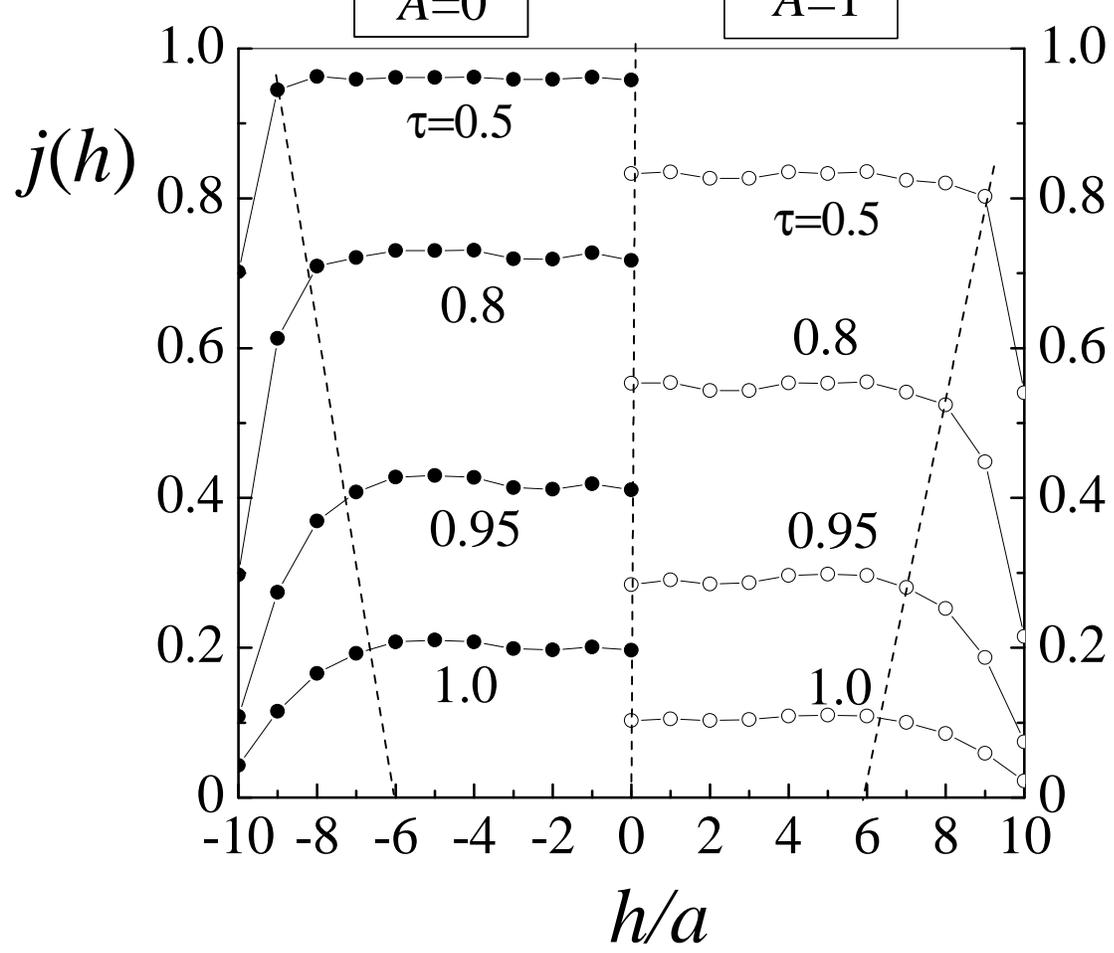
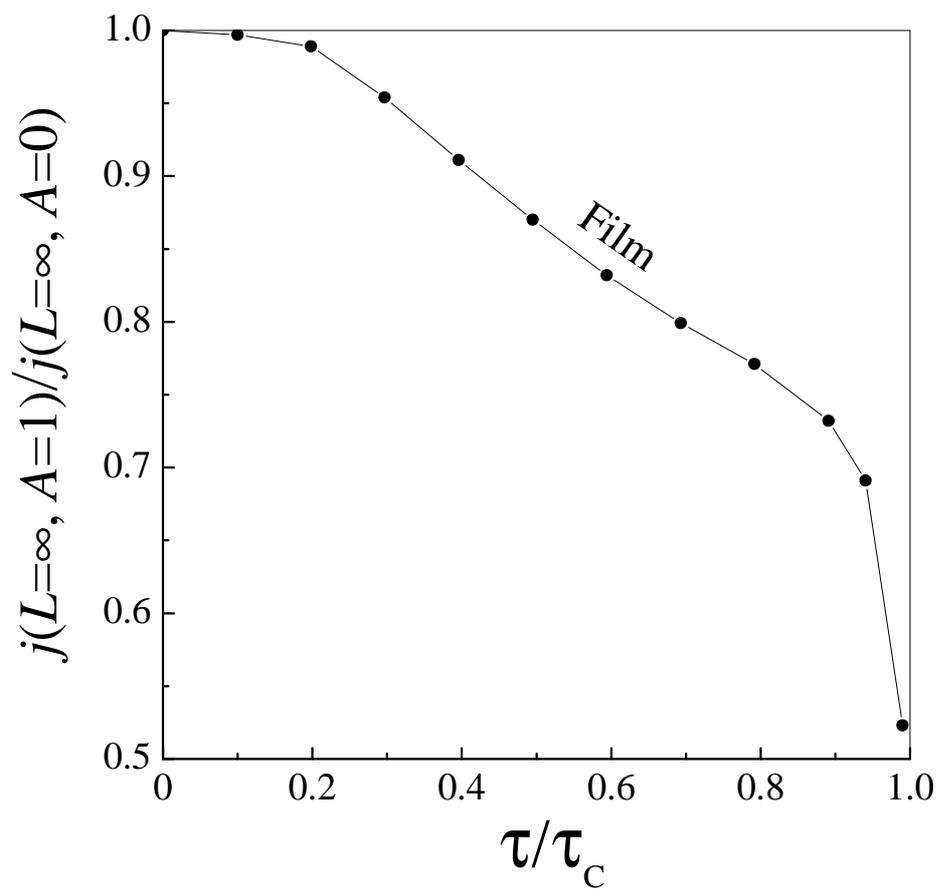
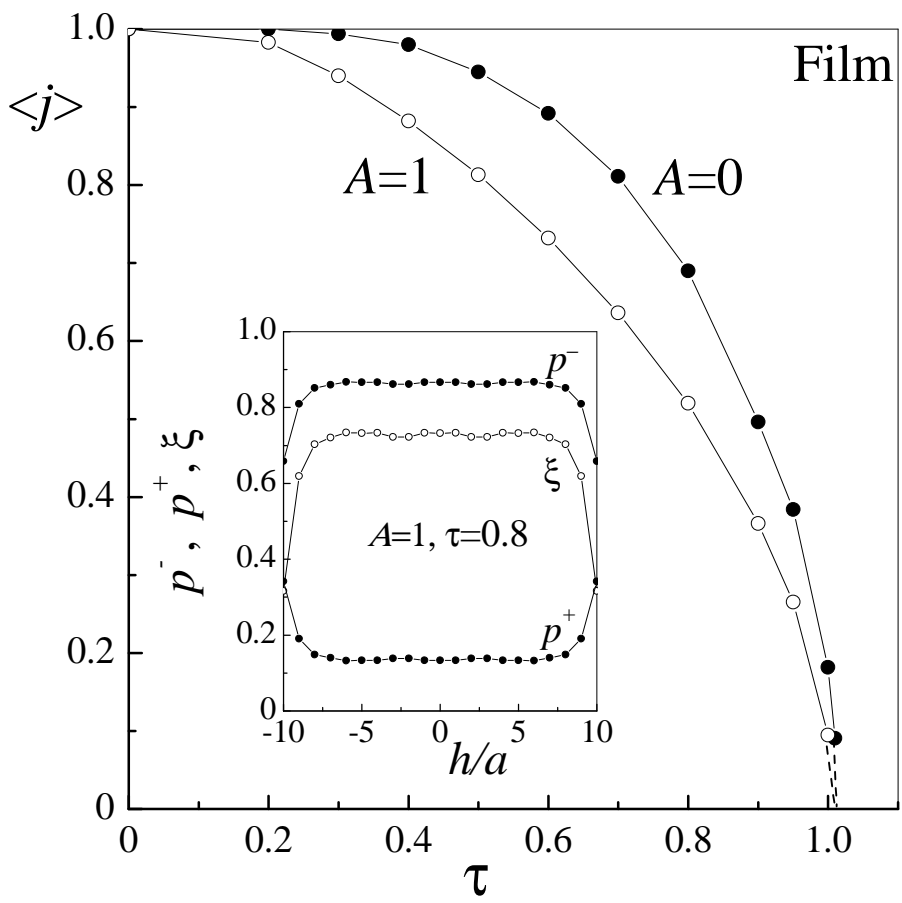
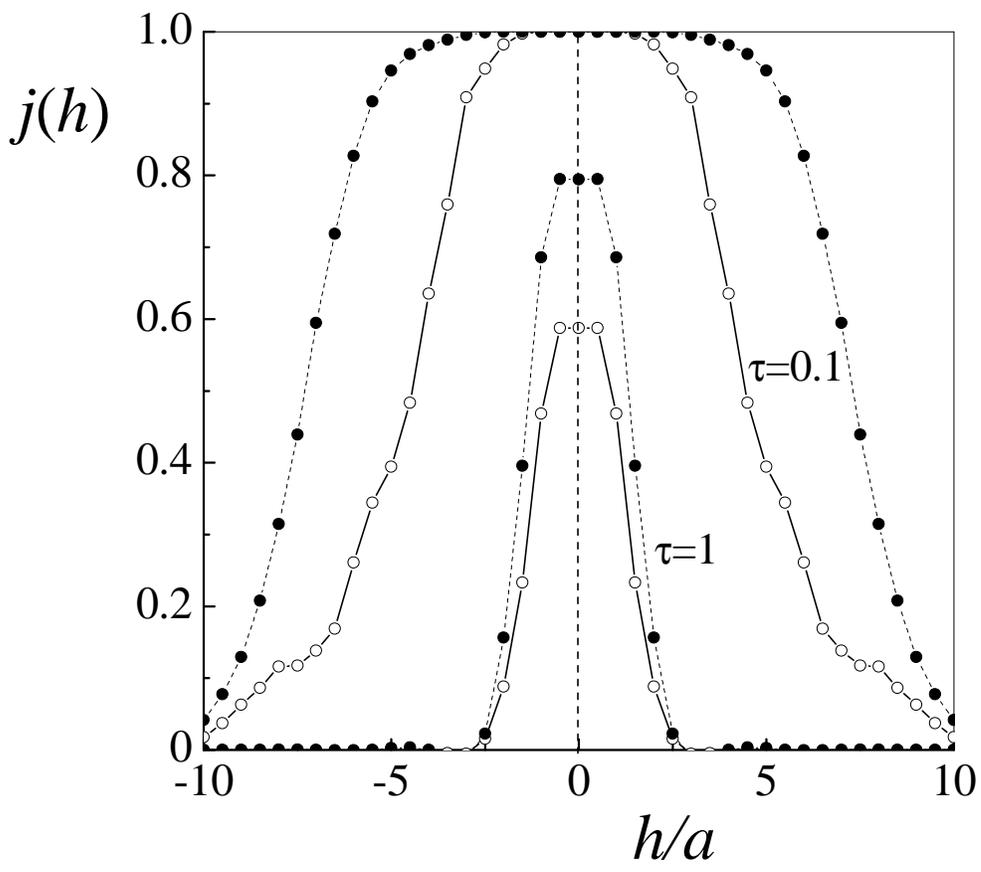
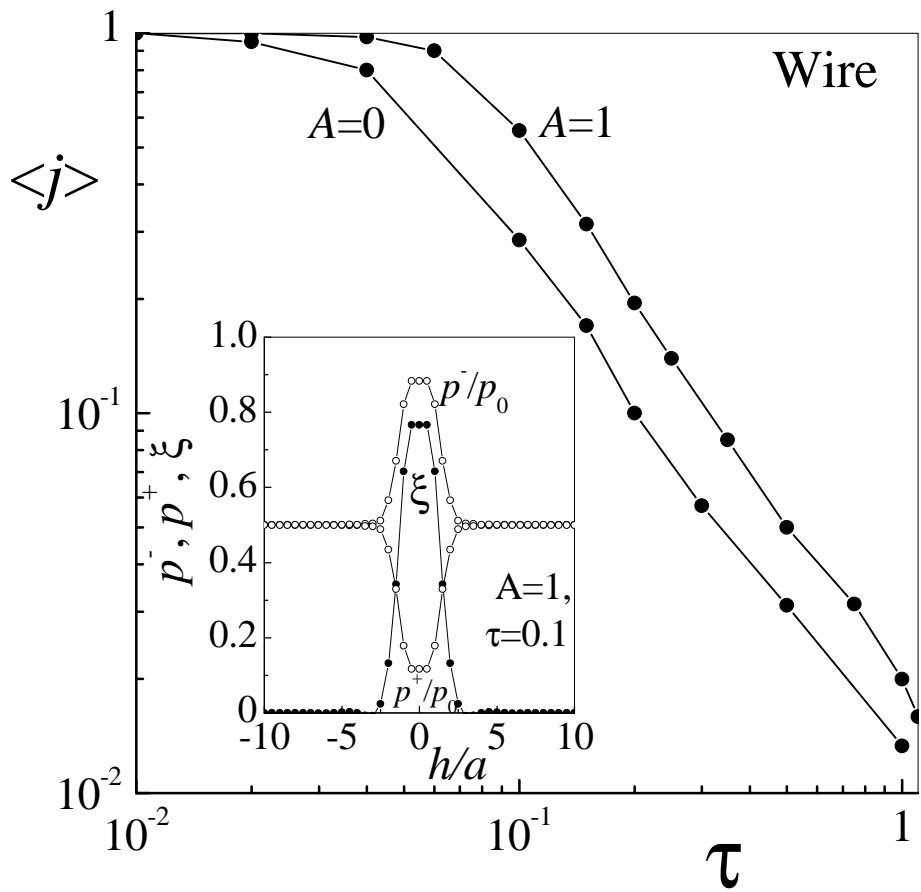
\end{document}